# Temperature- and field angular-dependent helical spin period characterized by magnetic dynamics in a chiral helimagnet MnNb$_3$S$_6$

Liyuan Li[1], Haotian Li[1], Kaiyuan Zhou[1], Yaoyu Gu[1], Qingwei Fu[1], Lina Chen[1,2,*], Lei Zhang[3,*], and R. H. Liu[1,*]

*1 National Laboratory of Solid State Microstructures, School of Physics and Collaborative Innovation Center of Advanced Microstructures, Nanjing University, Nanjing 210093, China*

*2 New Energy Technology Engineering Laboratory of Jiangsu Provence & School of Science, Nanjing University of Posts and Telecommunications, Nanjing 210023, China*

*3 Anhui Key Laboratory of Condensed Matter Physics at Extreme Conditions, High Magnetic Field Laboratory, Hefei Institutes of Physical Science, Chinese Academy of Sciences, Hefei 230031, China*

\* Corresponding authors:
Ronghua Liu, rhliu@nju.edu.cn,
Lina Chen, chenlina@njupt,edu.cn
Lei Zhang, zhanglei@hmfl.ac.cn

**Abstract:** The chiral magnets with topological spin textures provide a rare platform to explore topology and magnetism for potential application implementation. Here, we study the magnetic dynamics of several spin configurations on the monoaxial chiral magnetic crystal MnNb$_3$S$_6$ via broadband ferromagnetic resonance (FMR) technique at cryogenic temperature. In the high-field forced ferromagnetic state (FFM) regime, the obtained frequency $f$ vs. resonance field $H_{res}$ dispersion curve follows the well-known Kittel formula for a single FFM, while in the low-field chiral magnetic soliton lattice (CSL) regime, the dependence of $H_{res}$ on magnetic field angle can be well-described by our modified Kittel formula including the mixture of a helical spin segment and the FFM phase. Furthermore, compared to the sophisticated Lorentz micrograph technique, the observed magnetic dynamics corresponding to different spin configurations allow us to obtain temperature- and field-dependent proportion of helical spin texture and helical spin period ratio $L(H)/L(0)$ via our modified Kittel formula. Our results demonstrated that field- and temperature-dependent nontrivial magnetic structures and corresponding distinct spin dynamics in chiral magnets can be an alternative and efficient approach to uncovering and controlling nontrivial topological magnetic dynamics.

## 1. Introduction

Chiral helimagnets (CHM) possess nontrivial spin-textures with spiral or rotary alignment of spin moments, such as topological spin textures of magnetic skyrmions, which provide a platform to study the interesting topological physics and potential applications for spintronics[1-5]. $MnNb_3S_6$ and $CrNb_3S_6$ are typical chiral helimagnets with the same lattice structure[6,7], analogous electronic[8,9], and magnetic structures [10,11]. In the monoaxial chiral helimagnets[12], all spins are in the ab-plane and rotate at a definite angle along the c-axis due to the competition among magnetocrystalline anisotropy, the interlayer Heisenberg and Dzyaloshinskii-Moriya (DM) interactions. The DM interaction arises from losing the inversion center in the magnetic atoms sublattice. The Heisenberg interaction (coefficient $J$) prefers all spins forming collinear arrangements (ferromagnetic or antiferromagnetic alignments). In contrast, the chiral DM interaction (coefficient $D$) favors the non-collinear alignment of spins and facilitates chiral magnetic orders[12]. Thus, their competition generates a chiral helimagnet with a fixed spin helix period $L(0)$ determined by the ratio of two interactions $L(0) = \tan^{-1}(D/J)$[13,14].

However, under an external field $H$, the field-dependent Zeeman interaction will also compete with the above two magnetic interactions and can be used to achieve field-controllable spin textures[6,15,16]. Therefore, by tuning magnetic field or/and temperature, the chiral helimagnets can evolve from a CHM into a chiral magnetic soliton lattice (CSL) or forced ferromagnetic state (FFM) to achieve the minimal total energy in terms of the competition of several magnetic interactions[6,17,18]. Additionally, there exist several specific chiral spin textures deviating from the ideal helical state [Figure 1(a)], e.g., chiral conical phase (CCP)[19], tilted chiral magnetic soliton lattice (TCSL)[17], and CSL[13] depending on not only the amplitude but also the angle of the external field to the ab-plane for the strong easy-plane anisotropy in these monoaxial hexagonal crystals[17,20]. As the schematics are shown in Figure 1(a), the external field $H$ tilts the spin direction of the spin soliton lattice, modulates the spin helix period $L(0)$ to $L(H)$ at $H < H_c$, and finally turns it into the FFM regime at $H > H_c$.

The previous theoretical investigations of CHM[21,22] reported that the period of CSL can be described by the 1D chiral sine-Gordon model, which generally follows the formula $L(H_{in})/L(0) = 4K(k)E(k)/\pi^2$ [23-26]. where $K(k)$ and $E(k)$ are the elliptic integrals of the first and second kinds with modulus $k$ ($0 \leq k \leq 1$), respectively, and $H_{in}$ is the in-plane component of an external magnetic field. The elliptic modulus $k$ is given by $k/E(k) = (H_{in}/H_c)^{1/2}$ to minimize the CSL formation energy. The static and dynamic magnetic properties experiments confirm that the nontrivial spin configurations of these chiral helimagnets highly depend on the external magnetic field, dimensionality, and temperature[1,7,17,27-29]. Moreover, the Lorentz transmission electron microscopy also directly observed the temperature-dependent CSL state and its period in $CrNb_3S_6$[13,14,30,31]. However, for $MnNb_3S_6$ helimagnet with the same lattice structure as $CrNb_3S_6$, the Lorentz transmission electron microscopy measurement failed to identify the spatial period of CSL because $MnNb_3S_6$ has a much lower magnetic order temperature $T_c \sim 45$ K and the weak field modulation of the helix period[32]. Therefore, a high sensitivity technique that can catch the spiral period information of $MnNb_3S_6$ and its evolution with the external magnetic field and temperature is urgently needed.

Here, we perform the systematic ferromagnetic resonance experiment to investigate thoroughly the detailed dependence of magnetic dynamics corresponding to the nontrivial CSL in $MnNb_3S_6$ on the field magnitude, angle, and temperature. We find that chiral helimagnet $MnNb_3S_6$ exhibits a distinct field angular dependence of spin resonance in low-field nontrivial CSL from the uniform FMR in high-field FFM. Then, we propose a modified Kittel model considering partial helix spin textures, which can successfully describe the experimentally observed spin dynamics of the low-field nontrivial CSL at different temperatures. Moreover, the modified Kittel model also enables us to extract temperature- and field-dependent proportion of the helical spin texture and helical spin period ratio $L(H)/L(0)$, like the sophisticated Lorentz micrograph technique in most chiral helimagnets. The demonstrated method can generally be used as an alternative and easy-access approach to explore interesting magnetic dynamics not just in MnNb$_3$S$_6$ and other topologically nontrivial chiral magnets.

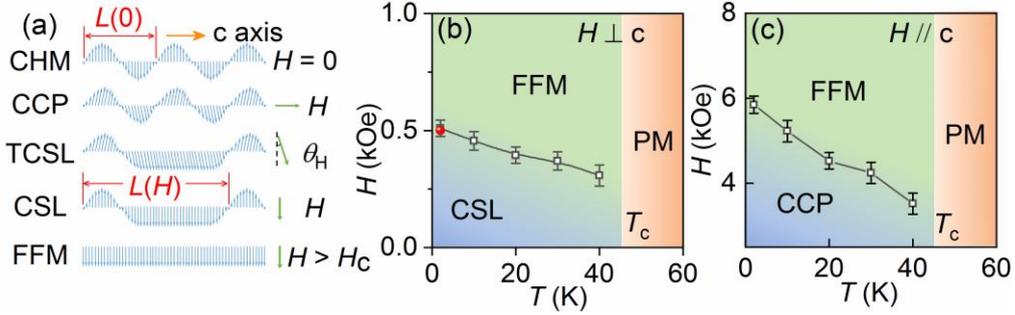

**Figure 1** Several nontrivial spin configurations and their phase diagrams of MnNb$_3$S$_6$. (a) Schematic of spin configuration of several magnetic orders in monoaxial chiral helimagnets under the external magnetic field: CHM state at $H = 0$, CCP, TCSL, CSL states at $0 < H < H_c$ and FFM state at $H > H_c$, respectively. The orange and green arrows represent the c-axis of the MnNb$_3$S$_6$ crystal and the direction of the magnetic field $H$, respectively. (b) - (c) Phase diagram of the specific magnetic orders in monoaxial chiral helimagnet MnNb$_3$S$_6$ crystal with $H$ // ab-plane (b) and $H$ // c axis (c). The boundaries among CSL (blue region) and FFM states (green region) were determined by critical field (squares) obtained from the quasi-static magnetization hysteresis loops. The critical field data (solid circle) reported by others is also shown in the phase diagram[7]. $T_c$ represents the Curie temperature 45 K of MnNb$_3$S$_6$, determined from the $M$ - $T$ curves. PM (orange region) represents paramagnetism.

Figure 1(b) and (c) show the phase diagram of spin textures in MnNb$_3$S$_6$ determined from the static field- and temperature-dependent magnetic susceptibility results with $H$ in the ab-plane and $H$ parallel to the c-axis of the single-crystal sample, respectively [see the Supporting Information (SI)[33]]. Note that the critical fields obtained by the static magnetization loop have some deviations from previous reports[11,34] due to different definition criteria and broad transition regions in $M(H)$ curves. More specifically, the critical field of the phase diagram in Figure 1(b) is slightly higher than in our previous reports[11]. One can find the detailed $M(H)$, $M(T)$ curves, and the definition criteria of the critical field in the SI [33]. The phase diagram of the studied chiral magnet MnNb$_3$S$_6$ shows two dominated spin configuration regions: a low-field CSL and a high-field FFM below its critical magnetic order temperature $T_c$ = 45 K, consistent with the previous reports [solid circle in phase diagram][7].

## 2. Experimental Section

The differential ferromagnetic resonance (FMR) spectroscopy is based on a coplanar waveguide (CPW), illustrated in Figure 2(a). A 1×1 mm square-shaped single-crystal MnNb$_3$S$_6$ with ~ 10 um thickness was fixed to the S-pole of the CPW by using the apiezon N-grease with high thermal conductivity and its c axis aligns along the z-axis [Figure 2(a)]. All cryogenic-temperature FMR spectra data were collected using a homemade differential FMR measurement system combining the lock-in technique and a closed-cycle G-M refrigerator-based cryostat. Static magnetic field $H$ can rotate in the y-z plane [Figure 2(a)] and be modulated with an amplitude of 1 - 2 Oe by a pair of secondary Helmholtz coils powered by an alternating current source with a low audio frequency of 129.99 Hz.

## 3. Results and Discussion

### 3.1 Spin resonance of the high-field FFM regime

Figure 2(b) shows the representative pseudocolor plot of normalized magnetic field-dependent FMR spectra obtained at excitation frequency $f$ varying from 5 to 20 GHz with 1 GHz steps, oblique field angle $\theta_H = 45°$ and cryogenic temperature $T = 4.5$ K. The inset of Fig. 2(b) exhibits a representative differential FMR spectrum with $f = 19$ GHz, which can be well fitted using a differential Lorentzian function. The characteristic dynamic properties, e.g., the resonance field $H_r$ and linewidth, can be extracted accurately from the fitting parameters of the experimental FMR spectra. As mentioned above, the external magnetic field can change the spin texture of MnNb$_3$S$_6$. For instance, the low-field CSL with a nontrivial topological property will be driven into the trivial FM state by an in-plane magnetic field $H_{in} \geq H_c \sim 0.51$ kOe at $T = 4.5$ K. Therefore, it is expected that the different dynamic properties corresponding to two distinct spin textures could be observed in our broadband FMR spectra.

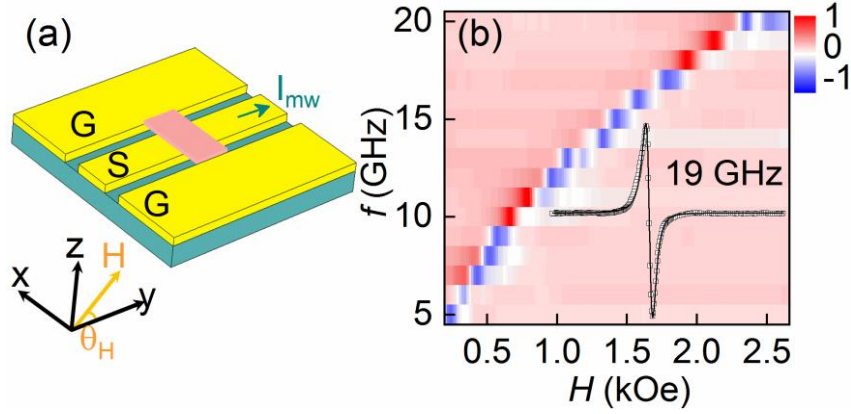

**Figure 2** Broadband differential FMR of flake crystal sample. (a) Schematic diagram of the high-sensitivity differential FMR experimental setup combining a coplanar waveguide technique. The green arrow represents the injection of microwave current. (b) Pseudocolor plot of the representative normalized magnetic field-dependent FMR spectra obtained at frequency $f$ between 5 and 20 GHz increased in 1 GHz steps, oblique field angle $\theta_H = 45°$ and temperature $T = 4.5$ K. The inset is a representative differential FMR spectrum (squares) obtained at $f = 19$ GHz and fitted by the differential Lorentzian function (solid black line).

To systematically explore the specific dynamics of chiral helimagnet $MnNb_3S_6$, we measured the broadband FMR spectra carefully at several different temperatures $T = 4.5$ K, 10 K, 30 K, and 45 K. Figure 3(a) - (d) show the frequency-dependent resonance field $H_{res}$ extracted by fitting experimental FMR spectra with a differential Lorentzian function[35,36]. For high-field range $H > H_c \sim 0.51$ kOe, we found that the dispersion curves of $f$ vs $H_{res}$ obtained at all four different temperatures can be well-fitted with the well-known Kittel formula as follows [see SI for specific derivation process]:

$$f = \gamma \sqrt{\begin{array}{c}(H_{res}\cos(\theta_M - \theta_H) + 4\pi M_{eff}\cos(2\theta_M)) \\ * (H_{res}\cos(\theta_M - \theta_H) - 4\pi M_{eff}\sin^2\theta_M)\end{array}} \quad (1)$$

where $\gamma = 2.9$ kOe/MHz is the gyromagnetic ratio, $4\pi M_{eff} = 4\pi M_s + H_k$ is the effective magnetization, $M_s$ is the saturation magnetization determined from static magnetization measurements, $H_k$ is the effective anisotropy field, the out-of-plane angle of the external field $\theta_H = 30°$ and magnetization $\theta_M$. For monoaxial chiral helimagnet $MnNb_3S_6$ with easy-plane anisotropy, the magnetocrystalline anisotropy constants $K_{u1}$, $K_{u2}$ are defined as $H_k = - [2(K_{u1}+2K_{u2})]/(\mu_0 M_s)$, where $K_{u2}$ can be neglected here because it is a fourth-order small item. Thus, $K_{u1}$ can be calculated by $K_{u1} = - (\mu_0 M_s H_k)/2$. The Kittel formula

(eq. (1)) can well fit the high-field dispersion relation, indicating that all spins have a uniform precession under the high-field range, consistent with discussed magnetic field-forced FM state at $H > H_c$ in the $H$ - $T$ phase diagram above [Figure 1(b)]. Note that, for $T = 45$ K, only $f$ vs. $H_{res}$ data in the high field range was used to be fitted because it exhibits a significant deviation at the low field range due to the strong spin fluctuation near its critical magnetic order temperature $T_c = 45$ K.

Futhermore, we can obtain temperature-dependent effective magnetization $M_{eff}$, out-of-plane angle of magnetization $\theta_M$, and magnetocrystalline anisotropy constant $K_{u1}$, which were together with the saturation magnetization $M_s$ measured by SQUID magnetometer shown in Figure 3(e) and (f). Analogous to $M_s$, $K_{u1}$ exhibits a monotonic decrease with increasing temperature and rapidly reduces to near zero while temperature approaches $T_c = 45$ K. Moreover, the temperature-dependent equilibrium position of magnetization $\theta_M$ shows that the magnetic moment is more accessible to follow external magnetic field $H$ due to the decrease of $H_k$ and demagnetized field with increasing temperature.

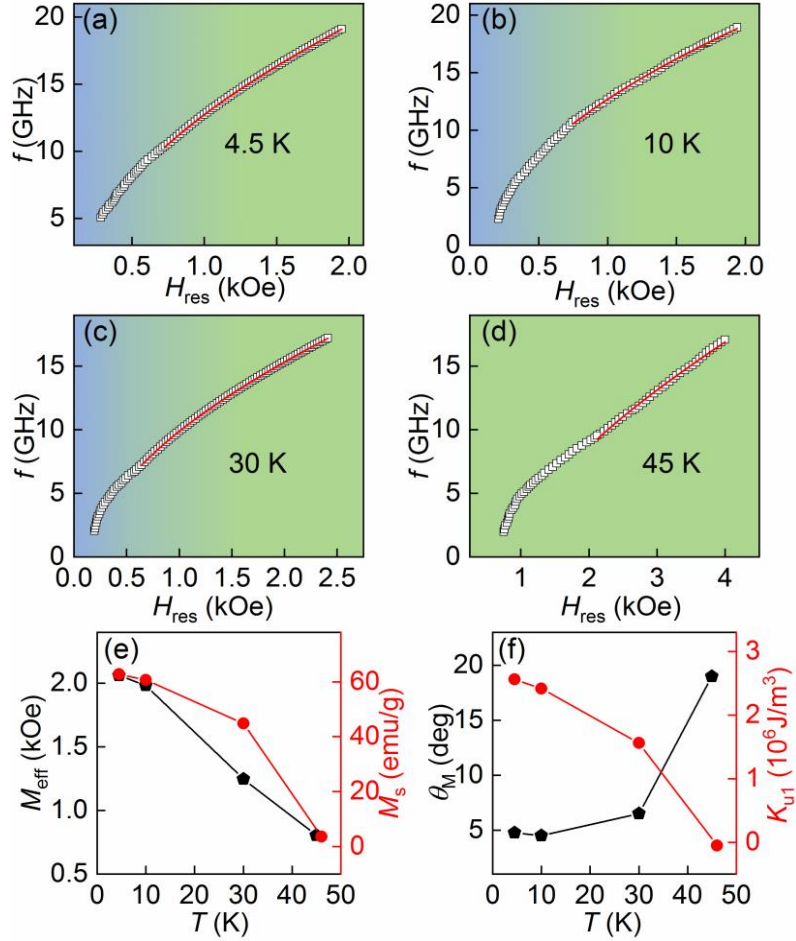

**Figure 3** Temperature dependence of the uniform FMR of the high-field FFM regime. (a) – (d) Symbols: $f$ vs $H_{res}$ experimental data obtained at oblique field angle $\theta_H$ = 30° and temperature $T$ = 4.5 K (a), 10 K (b), 30 K (c), and 45 K (d). The solid lines are the fitting results of the FMR data at the high-field range with the Kittel formula eq. (1). The regions with green and blue backgrounds represent the high-field FFM regime and low-field CSL regime, respectively. (e) – (f) Temperature-dependent effective magnetization $M_{eff}$ and saturation magnetization $M_s$ (e), the out-of-plane angle of magnetization $\theta_M$ and magnetocrystalline anisotropy constant $K_{u1}$ (f) were determined by fitting the FMR data using eq. (1) with best-fit parameters.

In addition to the discussed $f$ vs. $H_{res}$ dispersion relation above, the linewidth $\Delta H$, characterized by using the full width at half maximum (*FWHM*), can be used to analyze the Gilbert damping constant. *FWHM* is determined by fitting experimental FMR spectra with a differential Lorentzian function[35,36]. Figure 4(a) - (f) show the dependence of *FWHM* on the excitation frequency $f$ at several different temperatures $T$ = 4.5 K, 10 K, 20 K, 30 K, 40 K and 45 K with $H$ in the ab-plane( $\theta_H$ = 0°). For the high-field FFM regime ($H > H_c$), the relation of linewidth $\Delta H$ vs. $f$ can be well-fitted with the following formula $\Delta H = \Delta H_0 + \alpha f/\gamma$, where $\Delta H_0$ is the inhomogeneous linewidth broadening constant, α is the Gilbert damping factor. One can easily see that the linewidth obviously deviates from the linear fitting in the low-field range, which is caused by the emerging CSL phase in the low field, well consistent with the discussed $f$ vs. $H_{res}$ dispersion relation in Figure 3 above. The temperature dependence of the Gilbert damping constants α corresponding to high-field FFM regime [Inset of Figure 4(f)] shows a gradual enhancement with increasing temperature at the low-temperature range far below $T_c$ = 45 K, and then suddenly reaches 0.11 at 40 K from 0.05 at 30 K when the temperature approaches to $T_c$. The significant broadening of the linewidth near Curie temperature $T_c$ is related to the thermal effect-induced strong spin fluctuation.

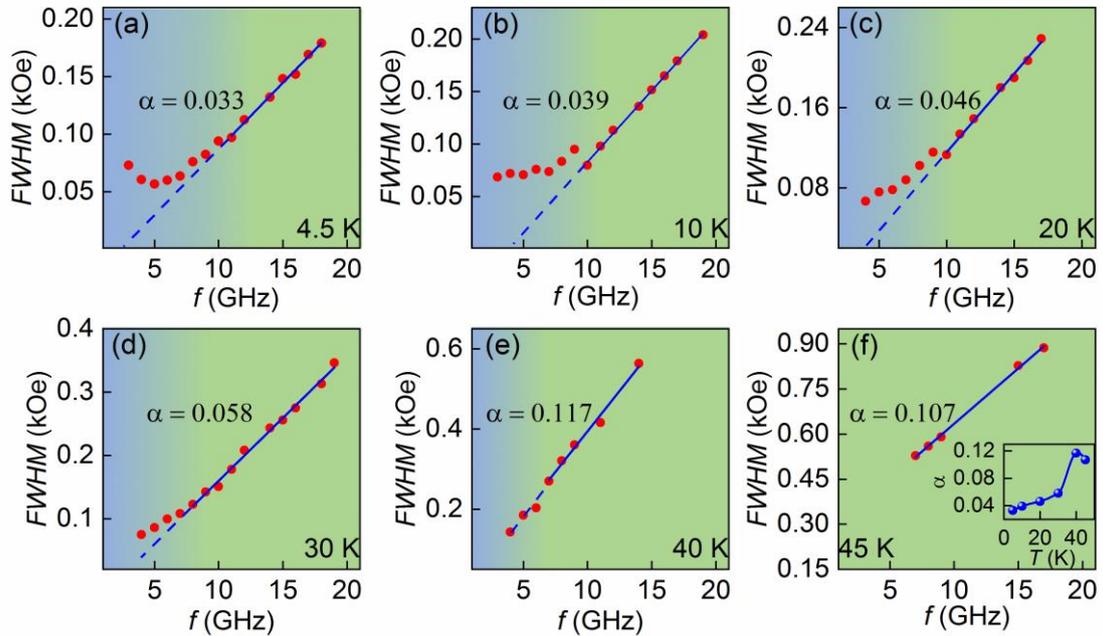

**Figure 4** Temperature- and field dependence of the FMR linewidth. (a) – (f) The FMR linewidth *FWHM* vs. *f* experimental data (symbols) obtained at in-plane field $H$ ($\theta_H$ = 0°), $T$ = 4.5 K (a), 10 K (b), 20 K (c), 30 K (d), 40 K (e) and 45 K (f). The linear lines are the fitting results of the *FWHM* data with $\Delta H = \Delta H_0 + \alpha f/\gamma$. The dashed lines are the extension of the linear fitting as guides to the eye. The regions with green and blue backgrounds represent the high-field FFM regime and low-field CSL regime, respectively. Inset in (f) Dependence of the Gilbert damping constant α on temperature determined by the linear fittings of the data in (a) – (f).

**3.2 Spin resonance of the low-field CSL regime**

Unlike the high-field FFM regime, the low-field CSL regime includes two magnetic structures, helical spin texture and FM phase. Compared to a single FM state, the mixture of helical spin segment and FM part in the CSL regime is expected to exhibit distinct magnetic dynamics due to the change of various magnetic interaction energies of the whole system. As discussed in Figure 3(a) and (d), the experimentally obtained dispersion results deviate significantly from the Kittel formula eq. (1) at the low-field range. Analogous to the unsaturated magnetic domain system[37], we derived a modified Kittel formula (eq. (2)) for this mixture of spin textures by reconsidering the total magnetic interaction energy of the system via setting the proportions of the helical spin segment and FM phase as $q$ and $p = 1- q$, respectively[see SI for specific derivation process] [33]. The modified FMR Kittel formula is given as follows:

$$f = \gamma \sqrt{\begin{aligned}&(q*Hsin\theta_M sin\theta_H + p*Hcos(\theta_M - \theta_H) + 4\pi M_{eff}\cos(2\theta_M))\\&*(q*Hsin\theta_M sin\theta_H + p*Hcos(\theta_M - \theta_H) - 4\pi M_{eff}\sin^2\theta_M)\end{aligned}} \quad (2)$$

where $1/q = L(H_{in})/L(0)$ can be proved strictly. Setting $q = 0$ in the modified Kittel model can be returned to the standard Kittel formula (eq. (1)) for the pure ferromagnetic state. As mentioned above, the proportion of helical spin texture $q$ depends significantly on in-plane field component $H_{in}$ [13,14]. Therefore, it is difficult to get a reliable fitting result about the experimentally obtained $f_{FMR}$ vs. $H_{res}$ dispersion relation in the low-field CSL regime. Because the spin helix period $L(H)$ (proportional to $1/q$) shows a significant in-plane magnetic field dependent divergency around critical field $H_c$ [13,14].

To further investigate spin dynamics in the low-field nontrivial CSL regime, we adopted out-of-plane angular-dependent FMR spectra. Because the critical field $H_c$ from CSL transferring to FFM is expected to be higher at large $\theta_H$ due to the demagnetization field and strong easy-plane magnetic anisotropy. More specifically, we quantitatively calculate the in-plane component of the resonance field obtained in the out-of-plane angular-dependent FMR spectra and find it only changes by 2.6% at $f = 6$ GHz, $T = 5$ K (see details in SI [33]), avoiding the in-plane field-induced significant modulation of $q$. Figure 5 shows the dependence of resonance field $H_{res}$ on out-of-plane angle $\theta_H$ from 0° to 90° with $T$ = 5 K at different resonant frequencies. Figure 5(a) - (h) show that the experimental angular-dependent $H_{res}$ results can be well fitted by the modified Kittel formula eq. (2) [see the detailed fitting process in the SI][33]. The non-zero helical spin proportion $q$ under low excitation frequency (less 8 GHz) indicates the existence of the CSL state in the studied oblique field range with a low in-plane component field $H_{in} < H_c$, consistent with the discussion of $f$ vs. $H_{res}$ curves at the in-plane field above. Figure 5(i) and (j) show the field dependence of the obtained fitting parameter $q$ and the helical spin period ratio $L(H_{in})/L(0)$. Helical spin proportion $q$ gradually decreases with increasing field and reaches zero corresponding to the disappearance of helical spin texture when the resonant field is above its critical field $H_c$ ~ 5.1 kOe at $T$ = 5 K, similar to the previously reported field-dependent spin helix period $L(H)$ of CSL state in chiral helimagnet $CrNb_3S_6$ by using the Lorentz micrograph technique[13,14]

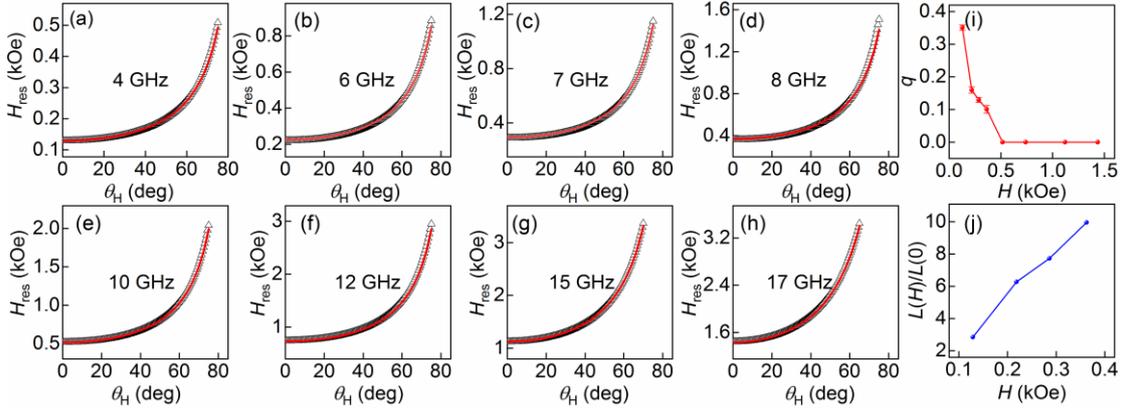

**Figure 5** Out-of-plane angular dependence of FMR spectra at 5 K. (a) – (f) The angular dependence of resonance field $H_{res}$ at $T$ = 5 K, $f_{ext}$ = 4 GHz (a), 6 GHz (b), 7 GHz (c), 8 GHz (d), 10 GHz (e), 12 GHz (f), 15 GHz (g), 17 GHz (h), respectively. The red solid lines are the results of fitting with the modified Kittel formula eq. (2) described in the main text. (i) – (j) Temperature dependence of the fitting parameter $q$ (i) and the helical spin period ratio $L(H_{in})/L(0)$ (j), respectively. The error bar of $q$ is defined in the SI [33].

To further investigate the temperature effect on the helical spin period of the low-field CSL regime, we also adopted out-of-plane angular-dependent FMR spectra at different temperatures below $T_c$. Figure 6 shows the dependence of $H_{res}$ on out-of-plane angle $\theta_H$ from 0° to 90° with $f$ = 6 GHz. Similarly, the obtained angular-dependent $H_{res}$ data can also be well fitted with the modified Kittel formula eq. (2) shown as red solid fitting curves in Figure 6(a) - (f). Figure 6(g) and (h) show the temperature dependence of the obtained fitting parameter $q$ and $L(H_{in})/L(0)$. $L(H_{in})/L(0)$ gradually increases with increasing temperature and reaches infinity when the resonant field is above its critical field $H_c$ at $T \geq 40$ K, also consistent with the previous report [13,14].

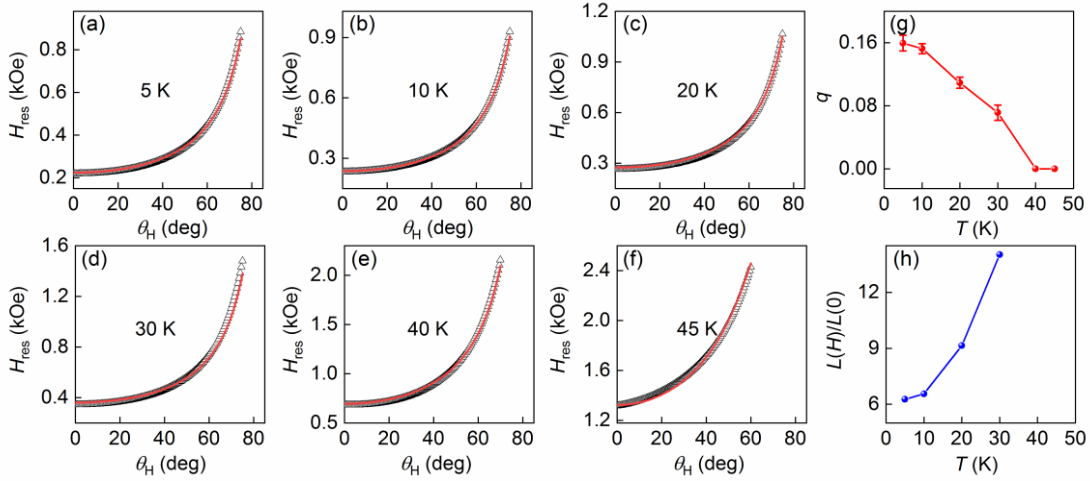

**Figure 6** Temperature-dependence of spin dynamics in the low-field CSL state. (a) – (f) The angular dependence of resonance field $H_{res}$ at $f_{ext} = 6$ GHz, $T = 5$ K (a), 10 K (b), 20 K (c), 30K (d), 40 K (e), 45 K (f), respectively. The red solid lines are the results of fitting with the modified Kittel formula eq. (2) described in the text. (g) – (h) Temperature dependence of the obtained fitting parameter $q$ (g) and the helical spin period ratio $L(H_{in})/L(0)$ (h), respectively.

### 3.3 Phase diagram determined by spin dynamics

In addition to the phase diagram consisting of the FFM and CSL state, as shown in Figure1(b), determined by the static magnetization characteristics, the dynamic analysis can also provide us with a detailed phase diagram of the CSL state. We quantitatively estimate the proportion of the helical spin texture $q$ (or helical period $L(H)$) from the angular-dependent dispersion relation of spin dynamics. We measure a series of out-of-plane angular-dependent FMR spectra with $f = 4, 7, 8, 10, 12, 15, 17$ GHz at different temperatures $T = 5, 10, 20, 30, 40, 45$ K (see the detail in SI [33]). After the analysis of dispersion relations as discussed above, we obtain the contour plot in terms of the component of the helical spin texture $q$ (equal to $L(0)/L(H)$) in the plane of temperature and in-plane field [Figure 7(a)], being overall consistent with the two-dimension phase diagram [Figure1(b)] determined by static magnetic susceptibility measurements. In the low-field CSL regime, the spin helix period $L(H)$ gradually increases with increasing the applied external in-plane magnetic field because an in-plane field can help to enhance the FM segment in CSL due to the Zeeman effect. Figure 7(b) shows that the helical spin proportion $q$ vs. normalized in-plane field $H/H_c$ curves obtained at different temperatures collapse into a single field dependence curve. Our results are consistent with the field dependence of $L(H)$ obtained by the Lorentz micrograph technique[13,14,31], confirming that analysis of out-of-plane field angular dependence of spin resonance using the modified Kittel formula can be regarded as another valid approach to probe the topological spin texture period in chiral magnets.

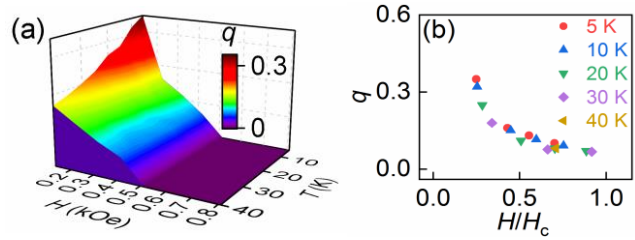

**Figure 7** Phase diagram determined by spin dynamics. (a) The contour plot in terms of the proportion of helical spin texture $q$ in the plane of temperature $T$ and in-plane magnetic field $H$. (b) Universal field scale dependence of $q$ obtained at different temperatures.

## 4. Conclusion

In summary, several specific spin textures and their distinct dynamics of the chiral helimagnet MnNb$_3$S$_6$ have been characterized detailly by field- and temperature-dependent static magnetization and broadband differential FMR spectroscopy. The high-field FFM follows the standard Kittel dispersion relation of a single domain FM state. In contrast, the low-field nontrivial CSL prefers the modified Kittel formula including the partial helix spin texture. Furthermore, like the sophisticated Lorentz micrograph technique, the modified Kittel model proposed in this work as an alternative and easy access approach enables us to extract the temperature- and field-dependent helical spin period ratio $L(H)/L(0)$ quantitatively from the angular-dependent FMR dispersion relation obtained at different temperatures. Our results find that the specific angular-dependent magnetic dynamics of nontrivial magnetic states proved in our work provide a vital clue to exploring interesting magnetic dynamics in other topologically nontrivial chiral magnets.

**Acknowledgments**

Project supported by the National Natural Science Foundation of China (Grant Nos. 11774150, 12074178, 12004171, 12074386, and 11874358), the Applied Basic Research Programs of Science and Technology Commission Foundation of Jiangsu Province, China (Grant No. BK20170627), the Open Research Fund of Jiangsu Provincial Key Laboratory for Nanotechnology, and the Scientific Foundation of Nanjing University of Posts and Telecommunications (NUPTSF) (Grant No. NY220164).

# Supporting Information

# Temperature- and field angular-dependent helical spin period characterized by magnetic dynamics in a chiral helimagnet MnNb$_3$S$_6$

Liyuan Li[1], Haotian Li[1], Kaiyuan Zhou[1], Yaoyu Gu[1], Qingwei Fu[1], Lina Chen[1,2,*], Lei Zhang[3,*], and R. H. Liu[1,*]

*1 National Laboratory of Solid State Microstructures, School of Physics and Collaborative Innovation Center of Advanced Microstructures, Nanjing University, Nanjing 210093, China*

*2 New Energy Technology Engineering Laboratory of Jiangsu Provence & School of Science, Nanjing University of Posts and Telecommunications, Nanjing 210023, China*

*3 Anhui Key Laboratory of Condensed Matter Physics at Extreme Conditions, High Magnetic Field Laboratory, Hefei Institutes of Physical Science, Chinese Academy of Sciences, Hefei 230031, China*

\* Corresponding authors:
Ronghua Liu, rhliu@nju.edu.cn,
Lina Chen, chenlina@njupt,edu.cn
Lei Zhang, zhanglei@hmfl.ac.cn

# Supporting Information

**1. The FMR Kittel formula for a single ferromagnet**

To systematically characterize the magnetization dynamics of the chiral helimagnet single-crystal MnNb$_3$S$_6$, we firstly studied the uniformed ferromagnetic resonance (FMR) of the field-forced ferromagnet (FFM) state under the high field range. For a ferromagnetic crystal with a single magnetic domain, the total free energy $E$ generally includes the magnetocrystalline anisotropic energy $E_k$, the external field Zeeman energy $E_H$, the demagnetized field energy $E_D$ and the exchange energy $E_{ex}$. In addition, since the uniform precession of magnetic moment in a single magnetic domain crystal, the exchange energy $E_{ex}$ only exists near the edge of the crystal and can be neglected. For the monoaxial chiral helimagnet MnNb$_3$S$_6$ with an easy-planar magnetocrystalline anisotropy, the whole free energy of the FFM state could be written as:

$$E = E_k + E_H + E_D$$
$$= K_0 + K_{u1}cos^2\theta - \mu_0 MH \cdot [sin\xi sin\theta + cos\xi cos\theta \cos(\eta - \varphi)] \quad (A1)$$
$$+ \frac{M^2}{8}N_y cos^2\theta + \frac{M^2}{8}N_z sin^2\theta$$

where $\theta = \theta_M$, $\xi = \theta_H$ with the same definition with the main text, $\varphi$ and $\eta$ are the azimuth angle of the magnetization $M$ and the external magnetic field $H$, while $N_y$ and $N_z$ are the demagnetization factor in the y-axis and z-axis directions, respectively. The first two terms at the right side of the above formula eq. (A1) are the magnetocrystalline anisotropic energy $E_k$; the third term is the external field Zeeman energy $E_H$; the last two terms are the demagnetized field energy $E_D$. The equilibrium position of the $M$ can be determined by $\partial E/\partial\theta = \partial E/\partial\varphi = 0$, which leads to

$$\begin{cases} -[K_{u1} + \frac{M^2}{8}(N_y - N_z)]sin2\theta - \mu_0 MHsin(\xi - \theta) = 0 \\ \eta = \varphi \end{cases} \quad (A2)$$

Therefore, the dispersion relation of the resonate frequency $f_{FMR}$ on the magnetic field $H$ is

$$f = \frac{\gamma}{\mu_0 M cos\theta}\sqrt{\left[\frac{\partial^2 E}{\partial\theta^2} \cdot \frac{\partial^2 E}{\partial\varphi^2} - \left(\frac{\partial^2 E}{\partial\theta\partial\varphi}\right)^2\right]}$$
$$= \gamma\sqrt{\begin{array}{c}(Hcos(\theta_M - \theta_H) + 4\pi M_{eff}\cos(2\theta_M)) \\ * (Hcos(\theta_M - \theta_H) - 4\pi M_{eff}sin^2\theta_M)\end{array}} \quad (A3)$$

where the effective magnetization $4\pi M_{eff}$ = $-2[K_{u1} + M^2(N_y - N_z)/8]/\mu_0 M$ = $(N_z - N_y)M/4\mu_0 - 2K_{u1}/\mu_0 M$. The effective magnetocrystalline anisotropy field $H_k$ also can be defined as $H_k = -(2K_{u1})/(\mu_0 M)$. Additionally, the effective magnetization is equal to $4\pi M_s + H_k$ because the studied magnetic crystal is a thin flake sample.

## 2. The modified Kittel formula for a chiral magnetic soliton lattice

To further study the magnetization dynamics of the chiral magnetic soliton lattice CSL (consisting of the mixture of a helical spin segment and an FM block), we performed the additional FMR experiments of the out-of-plane magnetic angular-dependent resonance field $H_{res}$ with a certain excitation frequency. To better analyze the dynamics of the nontrivial CSL, we rewrite the whole free energy by adding a term contributed from the partial helical spin segment in a similar way used in dealing with the unsaturated magnetic domain system[1]. Setting the proportions of the helical spin segment and the FM block in the CSL phase as $q$ and $p = 1 - q$, respectively, we can get

$$E_k = K_{u0} + K_{u1} cos^2\theta \tag{A4a}$$

$$E_H = -\mu_0 M H \cdot [q sin\xi sin\theta + p cos(\theta - \xi)] \tag{A4b}$$

$$E_D = \frac{pM^2}{8} N_y cos^2\theta + \frac{N_z}{2}(\frac{pM sin\theta}{2} + \frac{qM sin\theta}{2})^2 = \frac{pM^2}{8} N_y cos^2\theta + \frac{M^2}{8} N_z sin^2\theta \tag{A4c}$$

$$E = E_k + E_H + E_D \tag{A5}$$

Since only FMR acoustic mode (magnetization processions of different ferromagnet blocks is in-phase) was observed in the FMR experiments, the internal magnetic field between the ferromagnet segments (or the effective bias field caused by the helical spin segment) arise from the nonlinear spin configuration (the interlayer Heisenberg and Dzyaloshinskii-Moriya interactions) should keep a constant. Therefore, the interlayer exchange interaction will do not contribute to the dispersion relation of the experimentally observed FMR acoustic mode. Following the same procedure mentioned in the above part, the dispersion relation of $f_{FMR}$ vs. $H$ is given by:

$$f = \gamma \sqrt{\begin{array}{c}(q * H sin\theta_M sin\theta_H + p * H cos(\theta_M - \theta_H) + 4\pi M_{eff} cos(2\theta_M)) \\ * (q * H sin\theta_M sin\theta_H + p * H cos(\theta_M - \theta_H) - 4\pi M_{eff} sin^2\theta_M)\end{array}} \tag{A6}$$

Where $4\pi M_{eff} = -2[K_{u1} + M^2(pN_y - N_z)/8]/\mu_0 M = (N_z - pN_y)M/4\mu_0 - 2K_{u1}/\mu_0 M$ is the same as the effective magnetization $4\pi M_{eff}$ in the high-field forced FM regime above because $N_y$ is 0 for a thin flake sample. Therefore, in fitting the spin-dynamic data obtained in the low-field CSL regime, we used the effective magnetization $4\pi M_{eff}$ determined from the high-field forced FM regime. Based on the above modulated FMR Kittle formula, we can obtain the proportions of the helical spin segment $q$ and the FM block $p=1-q$ in the studied CSL phase from the experimental angular-dependent resonance frequency. Furthermore, we can get the helix period information of the nontrivial chiral magnetic soliton lattice through $1/q = L(H_{in})/L(0)$ and its temperature dependence.

**3. Static magnetization measurement**

We charactered the temperature and field-dependent static magnetization of the single crystal MnNb$_3$S$_6$ by superconducting quantum interference device (SQUID) magnetometer with $H$ in the ab-plane and $H$ parallel to the c-axis of the single-crystal sample, respectively. The phase diagram with $H$ in the ab-plane (and $H$ parallel to the c-axis) [Figure 1(b)] (and [Figure 1(c)]) of spin textures in MnNb$_3$S$_6$ was determined by the field- [Figure S1(a) and (b)] (and [Figure S1(e)]) and temperature-dependent magnetic susceptibility [Figure S1(c)] (and [Figure S1(f)]) curves with the applied field $H$ in the ab-plane (and $H$ parallel to the c-axis) of the sample. The critical field $H_c$ of the transition from the CSL state to the FFM state was defined where the magnetization $M$ approaches its saturation in the $M$-$H$ curves [top inset in Figure S1(a)] and the first-order differentiation d$M$/d$H$ begins to deviate from the level line at the dc field $H_{dc}$ [bottom inset in Figure S1(b)] [2]. The field cooling (FC) and zero-field cooling (ZFC) curves with $H$ = 40 Oe applying in the ab-plane and parallel to the c-axis are shown in Figure S1(c) and Figure S1(f), respectively. $M$ ($T$) curves show the Curie temperature $T_c \sim$ 45 K of MnNb$_3$S$_6$. An anomalous sharp peak is observed in the ZFC curve, consistent with the characteristic peak of chiral helimagnets and antiferromagnets. All behaviors are well consistent with the previously reported results of the chiral helimagnets[3]. Figure S1(d) shows the out-of-plane angular-dependent magnetization with the external field $H_{ext}$ rotated from the ab-plane to the c-axis at 2K, 40 K, and 60 K.

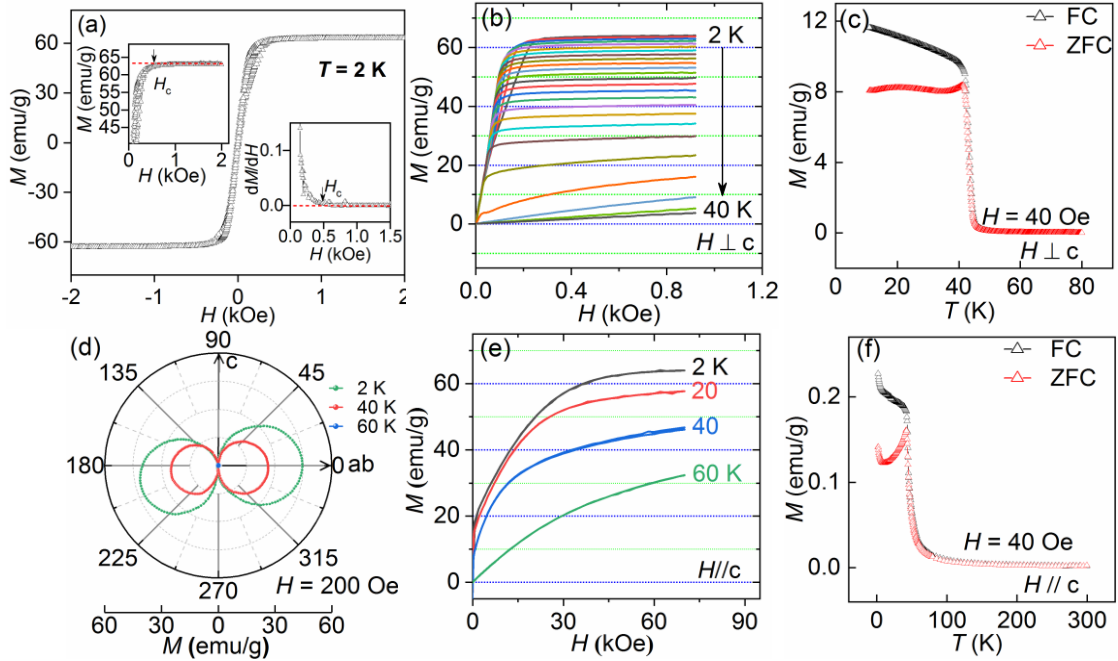

Figure S1. (a) Magnetization curves measured at dc field parallel to ab-plane of $MnNb_3S_6$ at temperature $T = 2$ K. The black arrow in the top inset marks the critical field $H_c$ of the transition from the CSL state to the FFM state. The bottom inset is the first-order differentiation $dM/dH$, which also can determine the critical field $H_c$, marked by the black arrow. (b) $M$-$H$ curves obtained with $H$ in the ab-plane at $T$ between 2 and 40 K increased in 2 K steps. (c) FC and ZFC curves with $H = 40$ Oe in the ab-plane. (d) Magnetization as a function of the out-of-plane angle measured at $H = 200$ Oe, $T = 2$ K, 40 K, 60 K. (e) $M$-$H$ curves with $H$ parallel to the c-axis at $T = 2$ K, 20 K, 40 K, 60 K. (f) FC and ZFC curves with $H = 40$ Oe parallel to the c-axis.

### 4. Analysis of out-of-plane angular-dependent FMR spectra

We systematically performed and analyzed the out-of-plane field angular-dependent FMR spectra with the excitation frequency $f = 4, 7, 6, 8, 10, 12, 15, 17$ GHz at different temperatures $T = 5, 10, 20, 30, 40, 45$ K. Based on the 1D chiral sine-Gordon model $L(H_{in})/L(0) = 4K(k)E(k)/\pi^2$ in the main text, we can quantitatively estimate the spiral order proportional $q(\theta_H)$ with the out-of-plane angle at a fixed excitation frequency (the in-plane component of the resonance field $H_{res}$). Taking out-of-plane angular-dependent FMR spectra at $f = 6$ GHz, $T = 5$ K as an example, we found that the in-plane component of the resonance field only changes by 2.6% [Figure S2(a)]. Therefore, we adopted the out-of-plane angular-dependent FMR spectra at a fixed frequency can avoid a field-induced significant change of $q$ [Fig. S2(b)].

For the low-field CSL regime, we fit the angular-dependent $H_{res}$ with a fixed excitation frequency by using the modified Kittel formula with three free parameters: the effective magnetization $M_{eff}$, the helical portion $q$, and the magnetization angle $\theta_M$. However, we can independently determine the effective magnetization $M_{eff}$ by the high-field FFM state using the standard Kittel formula because $M_{eff}$ should be equal for these two states under different fields at the same temperature. Second, to further minimize the deviation of q($\theta_H$) caused by the slight change of the in-plane component $H_{in}$ of the resonance field, we included the empirical relation of q($\theta_H$) using the 1D chiral sine-Gordon model $L(H_{in})/L(0) = 4K(k)E(k)/\pi^2$ in fitting the angular dependence of FMR results, The following Fig. S2(b) shows that the spiral order proportional q only has less 9 % change from $\theta_H = 0$ to 90º. Although the $H_{res}$ changes dramatically for the out-of-plane angular-dependent FMR spectra, the helical proportion $q$ or $L(H)$ still keeps almost no change (< 9 %) under the series out-of-plane resonance field $H_{res}$ with a constant excitation frequency (6 GHz) because the in-plane component of $H_{res}$ did not have significant change. Therefore, we use the average value $q$ as the spiral phase proportional under a fixed excitation frequency corresponding to a certain in-plane component of the resonance field. The error bar of $q$ was determined by the fitting deviation and slight change of $H_{in}$.

The equilibrium angle $\theta_M$ of the magnetic moment M was determined together by external magnetic field $H$, a sharp-induced demagnetized field, and an effective anisotropy field. We can obtain the $\theta_M$ vs. $\theta_H$ curve from the experimentally obtained $f$ vs $H$ dispersion curve at a certain $\theta_H$ with the FMR Kittel formula., as shown in Fig.S2(c).

# Supporting Information

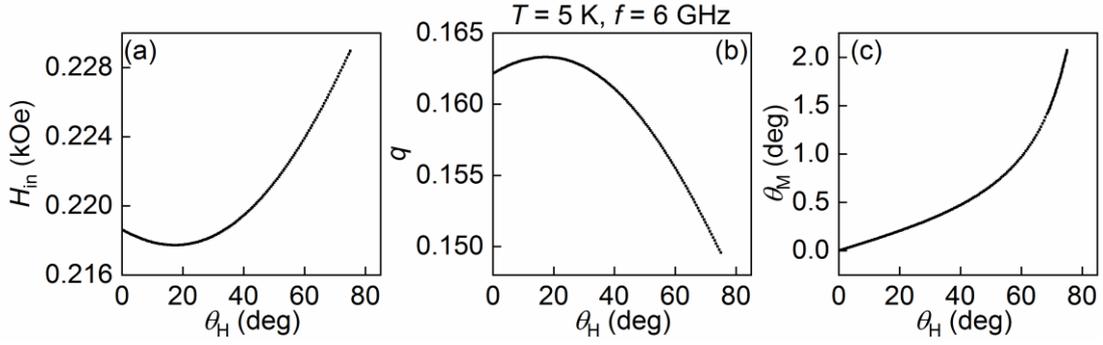

Figure S2. (a)-(c) the in-plane component of an external magnetic field $H_{in}$ (a), helical spin portion $q$ (b), and out-of-plane angle of magnetization $\theta_M$ (c) with the out-of-plane angle $\theta_H$ of the external field.

Furthermore, to intuitively prove the reliability and accuracy of the dynamic behavior analysis of the low-field CLS regime, we provided a more detailed analysis of out-of-plane angular-dependent FMR spectra at frequency $f$ = 4 GHz and $T$ = 5 K by using different helical proportions $q$ = 0, 0.1, 0.2, 0.3, 0.35 as the fitting parameters, as shown in Fig. 3S. The fitting curve with $q$ less 0.3 significantly deviates from the experimental data, indicating that the dynamic behaviors analysis of the CLS state using the modified Kittel model has good accuracy and sensitivity to estimate the helical portion $q$.

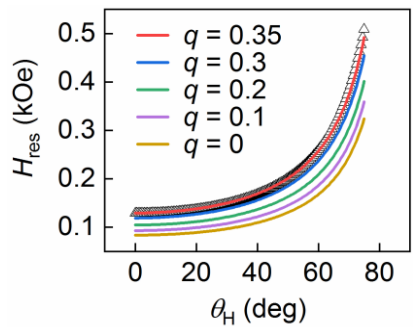

**Figure 3S** The angular dependence of resonance field $H_{res}$ at $f_{ext}$ = 4 GHz, $T$ = 5 K. The colored solid lines are the results of fitting with the modified Kittel formula with a fixing $q$ = 0.35, 0.3, 0.2, 0.1, 0, respectively.

Based on the discussion above, we fit the angular-dependent $H_{res}$ data obtained at $f$ < 8 GHz, $H$ < $H_c$ using the modified Kittel formula eq. (2) in the main text. The solid red curve represents the best fitting curve. The non-zero $q$ indicates the existence of the

# Supporting Information

CSL state in the studied oblique field range with a lower in-plane field component than the critical field $H_c$. The temperature dependence of the obtained fitting parameter $q$ and the helical spin period ratio $L(H_{in})/L(0)$ are summarized in Figs. S4(g) – S6(g) and Figs. S4(h) – S6(h), respectively. For $f$ > 10 GHz, the solid red curves in Figs. S7 – S10 represent the fitting curves with the Kittel formula eq. (1) (equal to $q = 0$ case in modified Kittel formula eq. (2)) instead of the modified Kittel formula eq. (2), indicating that the CSL state was destroyed or degraded by enhancement of magnetic field exceeding the critical field $H_c$.

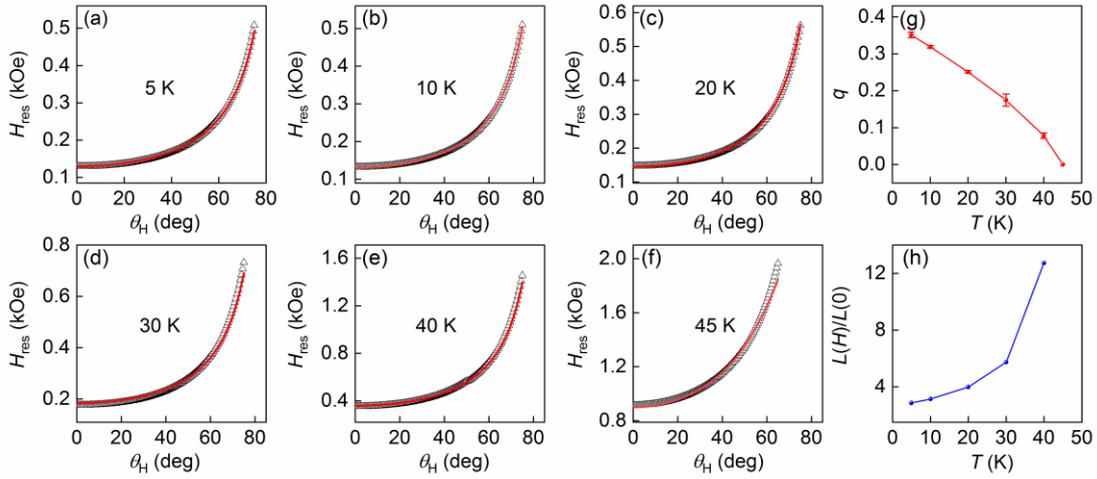

Figure S4. (a) – (f) The angular dependence of resonance field $H_{res}$ at $f_{ext}$ = 4 GHz, $T$ = 5 K (a), 10 K (b), 20 K (c), 30K (d), 40 K (e), 45 K (f), respectively. The solid red lines are the results of fitting with the modified Kittel formula eq. (2) described in the text. (g) – (h) The temperature dependence of the obtained fitting parameter $q$ (g) and the helical spin period ratio $L(H_{in})/L(0)$ (h), respectively.

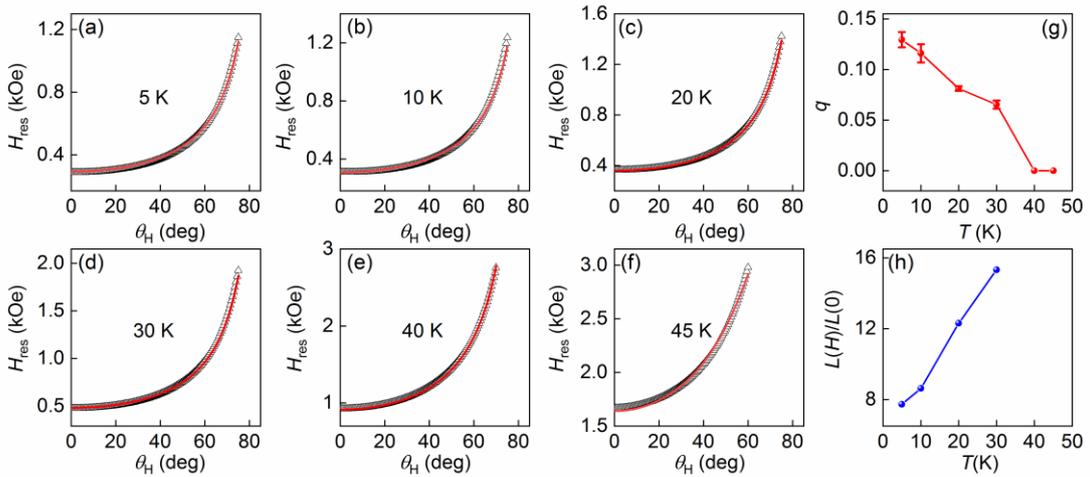

Figure S5. (a) – (f) The angular dependence of resonance field $H_{res}$ at $f_{ext}$ = 7 GHz, $T$ = 5

K (a), 10 K (b), 20 K (c), 30K (d), 40 K (e), 45 K (f), respectively. The solid red lines are the results of fitting with the modified Kittel formula eq. (2) described in the main text. (g) – (h) The temperature dependence of the obtained fitting parameter $q$ (g) and the helical spin period ratio $L(H_{in})/L(0)$ (h), respectively.

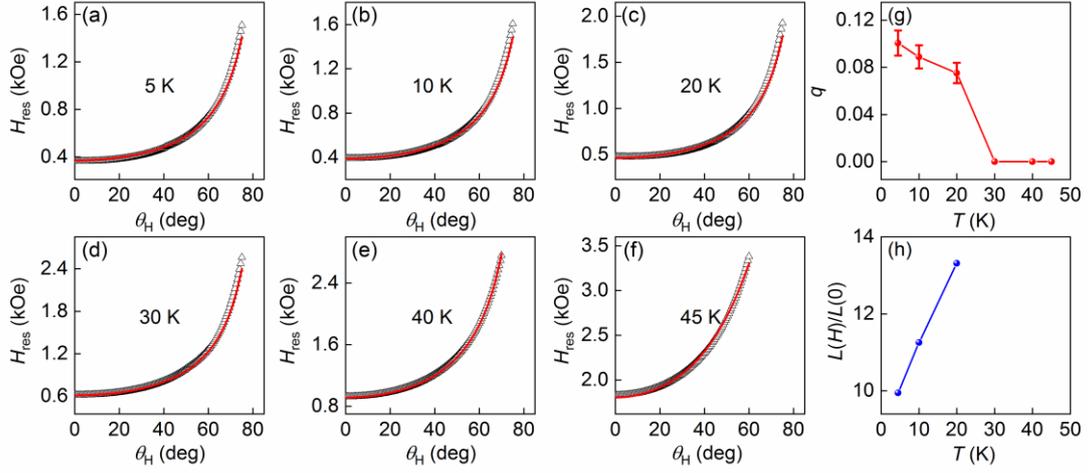

Figure S6. (a) – (f) The angular dependence of resonance field $H_{res}$ at $f_{ext}$ = 8 GHz, $T$ = 5 K (a), 10 K (b), 20 K (c), 30K (d), 40 K (e), 45 K (f), respectively. The solid red lines are the results of fitting with the modified Kittel formula eq. (2) described in the main text. (g) – (h) The temperature dependence of the obtained fitting parameter $q$ (g) and the helical spin period ratio $L(H_{in})/L(0)$ (h), respectively.

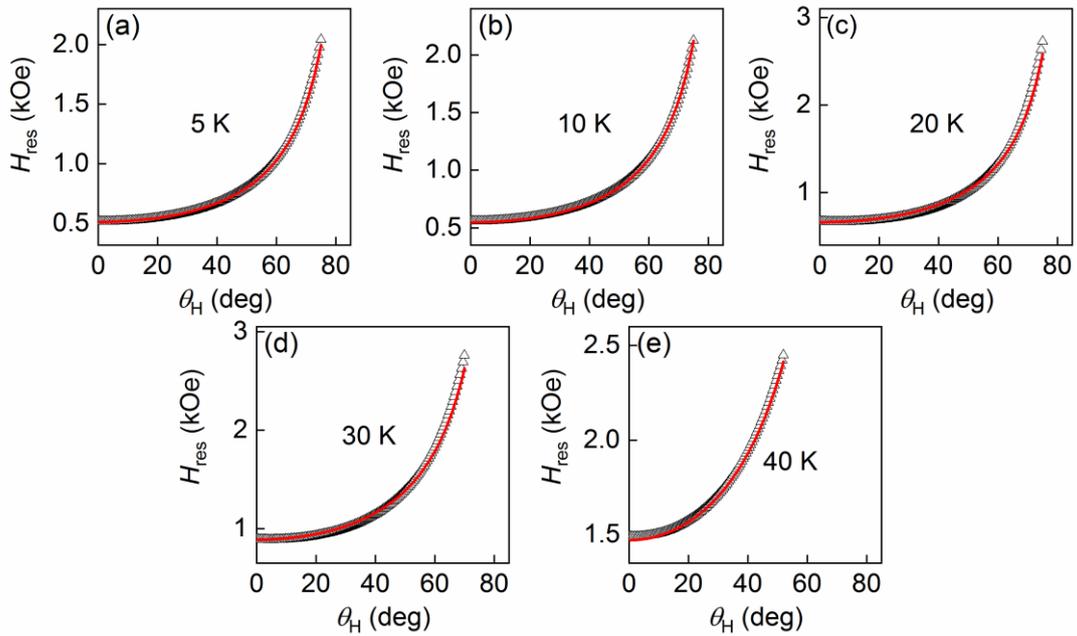

Figure S7. (a) – (e) The angular dependence of resonance field $H_{res}$ at $f_{ext}$ = 10 GHz, $T$ = 5 K (a), 10 K (b), 20 K (c), 30K (d), 40 K (e), respectively. The solid red lines are the results of fitting with the standard Kittel formula eq. (1) described in the main text.

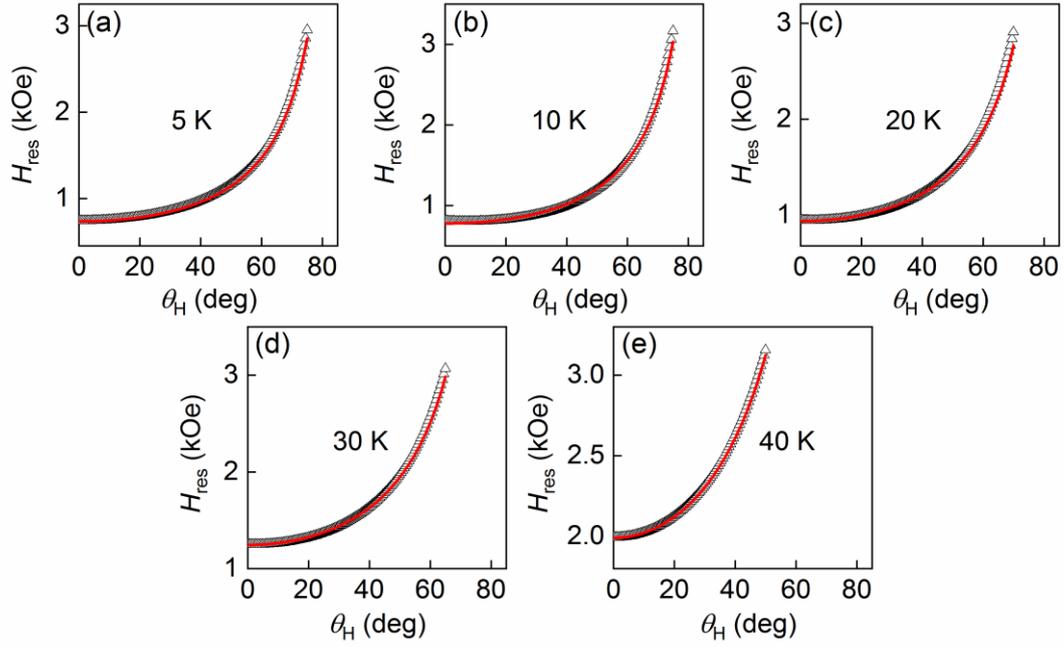

Figure S8. (a) – (e) The angular dependence of resonance field $H_{res}$ at $f_{ext}$ = 12 GHz, $T$ = 5 K (a), 10 K (b), 20 K (c), 30K (d), 40 K (e), respectively. The solid red lines are the results of fitting with the Kittel formula eq. (1).

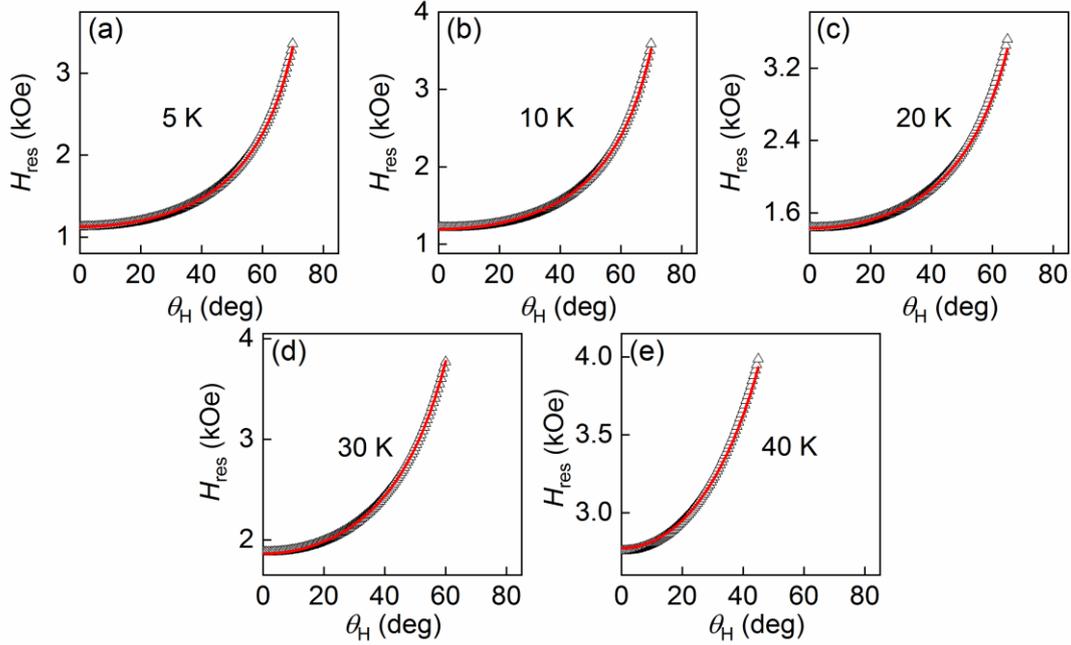

Figure S9. (a) – (e) The angular dependence of resonance field $H_{res}$ at $f_{ext}$ = 15 GHz, $T$ = 5 K (a), 10 K (b), 20 K (c), 30K (d), 40 K (e), respectively. The solid red lines are the results of fitting with the Kittel formula eq. (1).

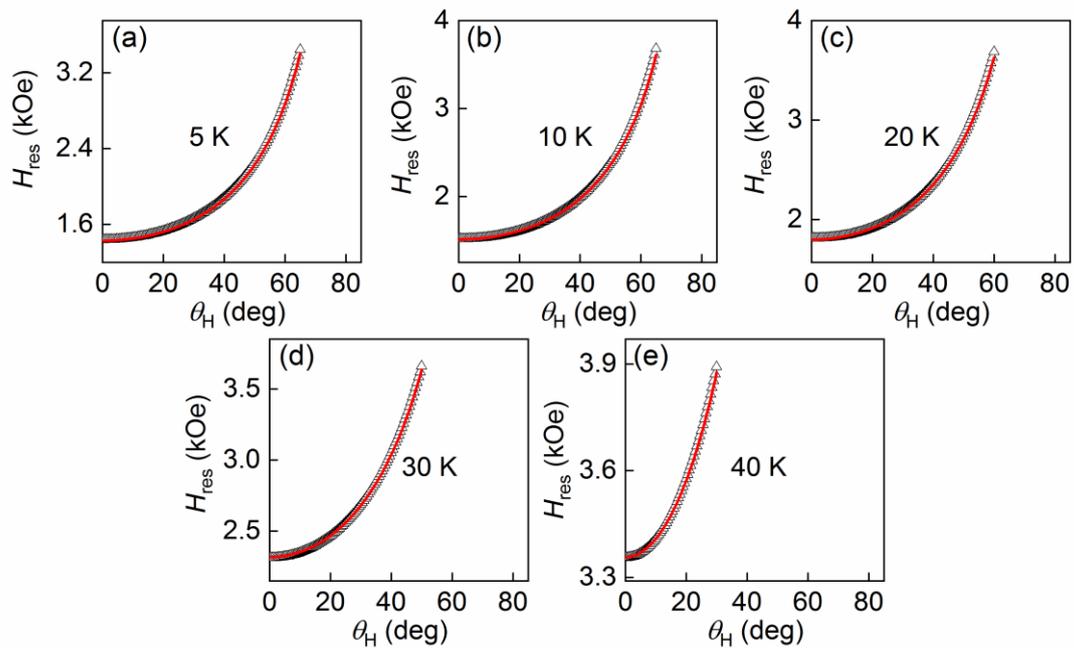

Figure S10. (a) – (e) The angular dependence of resonance field $H_{res}$ at $f_{ext}$ = 17 GHz, $T$ = 5 K (a), 10 K (b), 20 K (c), 30K (d), 40 K (e), respectively. The solid red lines are the results of fitting with the Kittel formula eq. (1).